\documentclass[preprint,superscriptaddress,a4paper]{revtex4}

\usepackage{graphicx}
\usepackage{amsmath}
\usepackage{lipsum}
\usepackage{float}

\usepackage{amsfonts}
\usepackage{xcolor}

\newcommand{\RNum}[1]{\uppercase\expandafter{\romannumeral #1\relax}}

\usepackage[justification=justified,singlelinecheck=false]{caption}
\begin{document}

\title{Stochastic analysis of bistability in coherent mixed feedback loops combining transcriptional and post-transcriptional regulations}

\author{Mor Nitzan}

\affiliation{Racah Institute of Physics, 
The Hebrew University, Jerusalem 91904, Israel}

\affiliation{Department of Microbiology and Molecular Genetics, 
IMRIC, Faculty of Medicine, 
The Hebrew University, Jerusalem 91120, Israel}

\author{Yishai Shimoni}

\affiliation{Racah Institute of Physics, 
The Hebrew University, Jerusalem 91904, Israel}

\affiliation{Department of Microbiology and Molecular Genetics, 
IMRIC, Faculty of Medicine, 
The Hebrew University, Jerusalem 91120, Israel}

\affiliation{Center for Computational Biology and Bioinformatics (C2B2), 
Columbia University, New York, NY, USA}

\author{Oded Rosolio}

\affiliation{Racah Institute of Physics, 
The Hebrew University, Jerusalem 91904, Israel}

\author{Hanah Margalit}

\affiliation{Department of Microbiology and Molecular Genetics, 
IMRIC, Faculty of Medicine, 
The Hebrew University, Jerusalem 91120, Israel}

\author{Ofer Biham}

\affiliation{Racah Institute of Physics, 
The Hebrew University, Jerusalem 91904, Israel}

\begin{abstract}

Mixed feedback loops combining transcriptional and 
post-transcriptional regulations are common 
in cellular regulatory networks.
They consist of two genes, encoding a transcription
factor and a small non-coding RNA (sRNA),
which mutually regulate each other's expression.
We present a theoretical and numerical study of 
coherent mixed feedback loops of this type, 
in which both regulations are negative.
Under suitable conditions, these feedback loops are expected to
exhibit bistability, namely two stable states, one dominated by the 
transcriptional repressor and the other dominated by the sRNA.
We use deterministic methods based on rate equation models, 
in order to identify the range of parameters in which bistability takes place.
However, the deterministic models do not account for the 
finite lifetimes of the bistable states and the spontaneous, 
fluctuation-driven transitions between them. 
Therefore, we use stochastic methods to calculate the average lifetimes of the two states.
It is found that these lifetimes strongly depend on rate coefficients
such as the transcription rates of the transcriptional
repressor and the sRNA.
In particular, we show that
the fraction of time the system spends in the
sRNA dominated state
follows a monotonically decreasing sigmoid function of the transcriptional repressor transcription rate.
The biological relevance of these results is discussed
in the context of such mixed feedback loops 
in {\it Escherichia coli}.

\end{abstract}



\maketitle


\section{Introduction}

\label{sec:introduction}

Recent studies of the interactions between molecules
in living cells revealed a complex 
interplay between regulatory interactions.
The regulatory mechanism that was most thoroughly investigated
is transcriptional regulation,
in which transcription factor (TF) proteins bind to specific promoter sites on the DNA
and regulate the transcription of downstream genes.
Recently, the significance of post-transcriptional regulation
by small non-coding RNA (sRNA) molecules has been recognized,
and is now known to play a major role in cellular processes 
\cite{Storz2005,Gottesman2005,Hershberg2003,Levine2007,Altuvia2000,Gottesman2011}.
It was suggested that this regulation mechanism 
would be energetically efficient,
since the sRNA molecules are relatively short and are not translated 
into proteins 
\cite{Altuvia2000}.
More recently, a quantitative analysis has
shown that post-transcriptional regulation by sRNAs 
provides fine tuning of the regulation strength \cite{Baker2012}
and is advantageous 
when fast responses to external stimuli are needed
\cite{Levine2007,Shimoni2007,Mehta2008}.

Regulatory interactions (such as the ones presented above) can be described
by a network  
in which genes and their products are represented by nodes,
while the interactions between them are represented by edges.
Analysis of such networks revealed 
structural modules or motifs, such as the auto-regulator and the feed-forward loop, which occur significantly more than randomly expected, and are expected to be of functional importance
\cite{Milo2002,Shenorr2002,Yeger-Lotem2003,Yeger-Lotem2004}.
Some of these motifs include only transcriptional regulation,
while others combine different layers of regulation
\cite{Shimoni2007,Yeger-Lotem2003,Yeger-Lotem2004,Zhang2005,Mandin2013}.

An important class of modules is the 
feedback loop, consisting of 
two genes, $a$ and $b$, that regulate each other's expression.
A well studied example of such module, in which both regulations
are at the transcriptional level, is the $\lambda$ switch in {\it E. coli}
\cite{Ptashne1992}.
Such transcriptional feedback loop,
referred to as the genetic toggle switch,
was constructed using methods of synthetic
biology and was shown to exhibit bistability
\cite{Gardner2000}.
Subsequent theoretical and numerical studies established
the conditions under which bistability takes place 
in such systems
\cite{Cherry2000,Kepler2001,Warren2004,Warren2005,Walczak2005,Lipshtat2006,Loinger2007}.

\begin{figure}

\includegraphics[width=7cm]{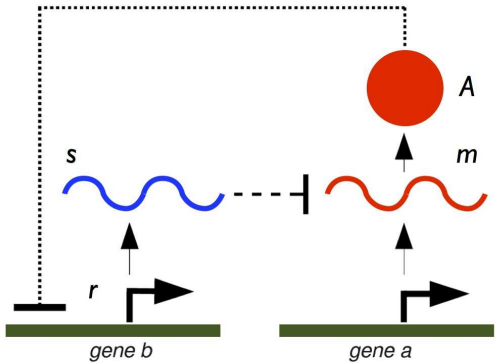}

\caption{
(Color online)
Schematic plot of the mixed feedback loop.
Gene {\it a} is transcribed into mRNAs ({\it m}), 
which are translated into transcriptional repressor proteins ({\it A}).
Gene {\it b} is transcribed into sRNAs ({\it s}), which are not translated
into proteins.
The {\it A} repressors negatively regulate the transcription of gene {\it b}
by binding to its promoter (the bound repressor is denoted by {\it r}).  
The sRNAs transcribed by gene {\it b} bind to the mRNAs of gene {\it a}
and inhibit their translation.
Truncated arrows represent negative regulation.
}

\label{fig1}

\end{figure}

In mixed feedback loops (MFLs), the two genes regulate each other 
using two different regulation mechanisms. A common form of MFLs involves a
gene {\it a} that expresses a TF and regulates gene {\it b}
via transcriptional regulation, while gene {\it b} transcribes a sRNA and regulates gene {\it a} via post-transcriptional regulation by 
sRNA-mRNA interaction.
In general, both the transcriptional regulator and the post-transcriptional regulator can act to either inhibit or activate their target.
MFLs in which both
regulations are negative (double-negative MFLs) belong to the
class of coherent feedback loops in which the number of negative
regulations is even. The positive-negative MFLs belong to the
class of incoherent feedback loops. In general, coherent feedback
loops tend to exhibit bistability while incoherent feedback loops
tend to exhibit oscillations, under appropriate parameter settings.
Schematic representation of the coherent MFL is shown in
Fig.~\ref{fig1}.

Integration of the transcriptional regulation network and the
network of sRNA-mRNA interactions in {\it E. coli}
has revealed that MFLs play important roles in various cellular contexts \cite{Shimoni2007}.
A textbook example of such a module is the
coherent MFL that consists of the TF Fur and the sRNA RyhB, involved in iron metabolism \cite{Masse2005,Masse2007,Vecercek2007}.
Another example of a coherent MFL in {\it E. coli} consists of the TF Lrp and the sRNA MicF, involved in cellular response to variation in nutrient availability \cite{Holmqvist2012}.


Other examples of MFLs involving noncoding RNAs (microRNAs) were found in the human regulatory network, playing a role in human granulopoiesis \cite{Fazi2005}, various cancers \cite{Fujita2008} and monocytic differentiation and maturation \cite{Fontanta2007}.
Further examples of MFLs 
were also found 
in {\it Drosophila melanogaster}
\cite{Varghese2008},
in {\it Vibrio harveyi} 
\cite{Tu2008},
and in {\it Caenorhabditis elegans}
\cite{Johnston2005}.
The identification of feedback loops and mixed-feedback loops
in the regulatory networks of various organisms 
hints to their important regulatory functions
\cite{Bela2008}.


The dynamic behavior of MFLs 
involving protein-protein interaction
was analyzed theoretically using
deterministic analysis in the framework of rate equation models
\cite{Francois2005}.
It was shown that within suitable ranges of parameters,
the double-negative (coherent) MFL exhibits bistability, 
while the positive-negative (incoherent)
MFL exhibits oscillations.
Similar results were recently obtained for MFLs involving 
sRNAs
\cite{Liu2011}
and
microRNAs
\cite{Zhou2012}.

Gene regulation processes are affected by fluctuations due 
to the stochastic nature of biochemical reactions and 
the fact that some of the molecules involved appear 
in low copy numbers. Therefore, in order to obtain a more
complete understanding of the dynamic behavior of coherent
MFLs and the emergence of bistability it is important to analyze
these systems using stochastic methods which take into account 
the discrete nature of the interacting RNAs and proteins as well as
the effects of fluctuations. 
Fluctuations in MFLs were previously characterized using
stochastic methods
\cite{Wang2011,Lloyd2013}.
However, the lifetimes of the two bistable states and their dependence on the model parameters have not been studied.

In this paper we present a comprehensive analysis of the dynamics
of a mixed coherent MFL involving transcriptional regulation and sRNA-mRNA interaction.
The analysis is done using a combination of deterministic and 
stochastic methods, enabling us to identify
the stable states of this system as well as the spontaneous, fluctuation-driven transitions
between them. 
We calculate the average
lifetimes of the two bistable states vs. parameters 
such as the transcription rates. 
As expected, we show that as the transcription rate of the mRNA, $g_m$, is increased, 
the average lifetime of the state dominated by the transcriptional
repressor, $\tau_a$,  increases, while the average lifetime of the
state dominated by the sRNA, $\tau_s$, decreases.
Thus, for small values of $g_m$ the system spends most of its time in
the sRNA dominated state, while for large values of $g_m$ it spends most
of the time in the state dominated by the transcriptional repressor. 
This means that in the two limits the domination times of the two regulators are biased towards one of the two bistable states.
The biological relevance of these observations is
discussed in the context of such MFLs apparent in {\it E. coli}.

The paper is organized as follows.
In Sec. 
\ref{sec:deterministic} 
we present a deterministic analysis of the MFL
and the results for the range of parameters in which
bistability appears.
In Sec. 
\ref{sec:stochastic} 
we present a stochastic analysis,
calculating the average lifetimes of the two bistable states 
vs. suitable parameters.
The results are summarized
and their biological relevance is discussed 
in Sec.
\ref{sec:summary}. 
In Appendix A
we present a detailed account of 
the experimental data we have used and
the considerations we have
made in order to determine the biologically relevant values of
the rate coefficients used in our model.
In Appendix B we extend the analysis of the bifurcation diagrams to a broader family of parameter variations and to the case in which the transcriptional repressor exhibits cooperative binding.


\section{Deterministic analysis}

\label{sec:deterministic}

Consider an MFL in which gene {\it a} encodes a transcription
factor and gene {\it b} encodes a small RNA.
Gene {\it a} negatively regulates gene {\it b} 
by transcriptional regulation, 
while gene {\it b} negatively regulates gene {\it a} 
post-transcriptionally via sRNA-mRNA interaction. 
In this system
gene {\it a} is transcribed into mRNA molecules, denoted by $m$,
which are translated into transcriptional repressor proteins, denoted by $A$. 
Gene {\it b} is transcribed into sRNA molecules, denoted by $s$. 
The transcriptional repressors $A$ negatively regulate
gene {\it b} by binding to its promoter site,
while the sRNAs negatively regulate gene {\it a} 
by binding to mRNA molecules, destabilizing them and
inhibiting their translation.

Here we describe the dynamics of a single MFL, namely one TF gene and one sRNA gene, using rate equations.
We denote the levels or copy numbers of 
the sRNA and mRNA molecules 
in the cell by
$s$ and $m$, respectively.
The level of the  sRNA-mRNA complex 
is denoted by $C$. 
The number of free $A$ proteins is denoted by $A$.
The number of $A$ proteins 
that are bound to the 
promoter site of gene {\it b} is denoted by $r$.
For simplicity, we consider the case in which the regulation 
is performed by a single copy of the bound repressor. 
In this case, $r$ takes values
in the range 
$0 \le r \le 1$.

The rate coefficients
$g_m$ and $g_s$
denote the transcription rates of genes 
{\it a} and {\it b},
respectively.
The translation rate of gene {\it a}, namely
the generation rate of $A$ proteins 
per copy of the mRNA molecule
is denoted by $g_A$.
The degradation rates 
of the sRNAs, mRNAs and the $A$ proteins
are denoted by 
$d_s$, $d_m$ and $d_A$, respectively. 
The binding rate of $A$ proteins 
to the promoter site of gene {\it b}
is denoted by 
$c_g$ 
and their dissociation rate from the promoter is denoted by
$u_g$.
The binding rate of sRNA and mRNA molecules to form a
complex is denoted by
$c_{ms}$. 
The sRNA-mRNA complex degrades at rate $d_c$, 
or dissociates into its 
sRNA and mRNA components at rate
$u_c$.
The processes taking place in the MFL and their rates are listed in Table \RNum{1}.

\begin{table}[H]
\caption{: The processes and respective rates in the MFL module.}
\begin{center}
  \begin{tabular}{| c  | c | c |}
    \hline
      & \textbf{Process} & \textbf{Rate} \\ \hline \hline
      1 &  $\o \ \to \  m$ & $g_m$  \\ \hline
      2 &  $m \ \to  \ \o$ & $d_m m$  \\ \hline
      3 &  $m + s \ \to \  C$ & $c_{ms} m s$  \\ \hline
      4 &  $C\ \to \  m + s$ & $u_c C$  \\ \hline
      5 &  $\o \ \to \  s$ & $g_s (1-r)$  \\ \hline
      6 &  $s \ \to  \ \o$ & $d_s s$  \\ \hline
      7 &  $\ m \ \to \  m + A \ $ & $g_A m$  \\ \hline
      8 &  $A \ \to \  \o$ & $d_A A$  \\ \hline
      9 &  $A \ \to \  r$ & $c_g A (1-r)$  \\ \hline
      10 &  $r \ \to \  A$ & $u_g r$  \\ \hline
      11 &  $C \ \to \  \o$ & $d_c C$  \\ \hline
  \end{tabular}
\end{center}
\label{table:reactions}
\end{table}

In Appendix \ref{sec:parameters} we consider the biologically relevant range
of values of each of the rate coefficients used in the analysis of 
the MFL. These values are determined on the basis of experimental results
and related considerations and interpretation.
For the calculations and simulations presented below we chose a default
value for each parameter, within the biologically relevant range.
These default parameter values, 
$g_m$=0.007, $g_s$=0.43, $g_A$=0.05, 
$d_m$=0.003, $d_s$=0.0008, $d_A$=0.001, 
$c_{ms}$=0.02, $c_g$=0.08 and $u_g$=0.01,
are used in all the Figures presented in this paper
(unless stated otherwise). 
All the parameters are in units of sec$^{-1}$.

The rate equations that describe this system take the form
\begin{subequations}
\begin{align}
\frac{dm}{dt} &= g_m - d_m m - c_{ms} m \cdot s + u_c C \\
\frac{ds}{dt} &= g_s (1-r) -d_s s - c_{ms} m \cdot s + u_c C \\
\frac{dA}{dt} &= g_A m - d_A A - c_g A (1-r) + u_g r \\
\frac{dr}{dt} &= c_g A(1-r) - u_g r \\
\frac{dC}{dt} &= c_{ms} m \cdot s - d_c C - u_c C, 
\end{align}

\label{eq:dns-mfl}

\end{subequations}

\noindent
where Eqs. 
(\ref{eq:dns-mfl}a)
and
(\ref{eq:dns-mfl}b)
account for the time dependent levels of 
the mRNAs and sRNAs, respectively. 
Each of these equations includes a transcription 
term and a degradation term. 
They also include binding terms, which describe the formation
rate of the sRNA-mRNA complex, and a term which represents the
dissociation of the complex.
The transcription term of $s$ includes
the factor $1-r$, which accounts for the fact that transcription takes
place only when there is no $A$ repressor bound to the $b$ promoter.
Eq. 
(\ref{eq:dns-mfl}c)
accounts for the time dependent level of the
$A$ protein, and includes translation and degradation
terms as well as terms describing the binding/unbinding 
to/from the $b$ promoter. 
Eq.
(\ref{eq:dns-mfl}d)
accounts for the level of $A$ proteins 
which are bound to the {\it b} promoter.
Eq.
(\ref{eq:dns-mfl}e)
accounts for the level of the 
sRNA-mRNA complex.

The rate equations can be solved by direct numerical integration.
For fixed values of the parameters and for a given choice of the initial
conditions, the system tends to converge to a steady state.
Coherent feedback loops such as the MFL 
tend to exhibit bistabily within a suitable range of parameters. 
In such cases, the steady state to which the rate equations converge
depends on the initial conditions.
Within the rate equation model, once the system converges to one
of the bistable states, it remains there and does not switch to the
other state.

Under steady state conditions (or in the limit in which
the formation and dissociation processes of the sRNA-mRNA complex are
fast) the effect of the dissociation process 
on the RNA and protein levels
can be accounted for by a suitable adjustment
of the binding rate coefficient $c_{ms}$.
Therefore, the dissociation process is expected to be of secondary importance 
and does not affect the essential properties of the MFL.

In the analysis presented below 
it is assumed, for simplicity, that the dissociation rate
$u_c = 0$, namely,
once an sRNA-mRNA complex is formed, it goes to degradation
rather than dissociate into its sRNA and mRNA components. 
Under this assumption, the level of the sRNA-mRNA complex, 
$C$, has no effect on the levels of other components in the MFL. 
Therefore, 
the set of four equations
(\ref{eq:dns-mfl}a)-(\ref{eq:dns-mfl}d)
can be integrated numerically or solved separately from
Eq. (\ref{eq:dns-mfl}e).

Under steady state conditions, the time derivatives on the
left hand side of 
Eqs.
(\ref{eq:dns-mfl})
vanish, and the rate equations are reduced to a set of coupled
algebraic equations.
These equations can be transformed into a single cubic 
equation of the form  
\begin{equation}
\left( m - \frac{g_m}{d_m} \right)
\left( m + \frac{d_s}{c_{ms}} \right)
\left( m + \frac{u_g}{c_g} \frac{d_A}{g_A} \right)
+ \frac{u_g}{c_g} \frac{d_A}{g_A} \frac{g_s}{d_m} m = 0
\label{eq:dns-mfl-steady}
\end{equation}

\noindent

For convenience we define the following dimensionless parameters:
$M = g_m/d_m$,
$D = d_s/c_{ms}$,
$K = u_g d_A/(c_g g_A)$,
and
$S = g_s/d_m$.
The parameter $M$ represents the average number of mRNA molecules
in the cell in the case that they are not regulated by sRNAs.
The parameter $D$ is
the ratio between the probabilites that a single sRNA will
degrade or bind to a single mRNA target.
Therefore, $D$ tends to decrease as the
strength of the regulation by sRNAs increases.
The parameter $K$ is inversely proportional to the
level of $A$ proteins (when unregulated) and to their binding
affinity to the $b$ promoter. 
Therefore, $K$ decreases as the transcriptional regulation
of gene $b$ becomes stronger.
The parameter $S$ represents the number of sRNA molecules
which are transcribed during the average
lifetime of an mRNA molecule.
Using these parameters, we obtain a cubic equation for $m$ of the form
\begin{equation}
m^3 + a_2 m^2 + a_1 m + a_0 = 0,
\label{eq:cubic_m}
\end{equation}

\noindent
where
$a_0 = -K D M$,
$a_1 = K (D + S - M) - M D$
and
$a_2 = D + K - M$.
Depending on the values of the parameters, 
this equation may have either one or three
real and positive solutions.
In the first case, the system exhibits a single steady state.
In the second case, it exhibits bistability, namely two stable 
steady states, while the third solution is unstable.

To analyze the existence and stability of the 
steady states of the MFL, it is useful to consider the 
bifurcation diagrams, presenting the steady state levels of 
$A$ and $s$ as a function of different parameters of 
the model. The stability of the solutions can be determined 
from the Jacobian of the set of rate equations  
and its eigenvalues.

\begin{figure}

\includegraphics[width=0.8\textwidth]{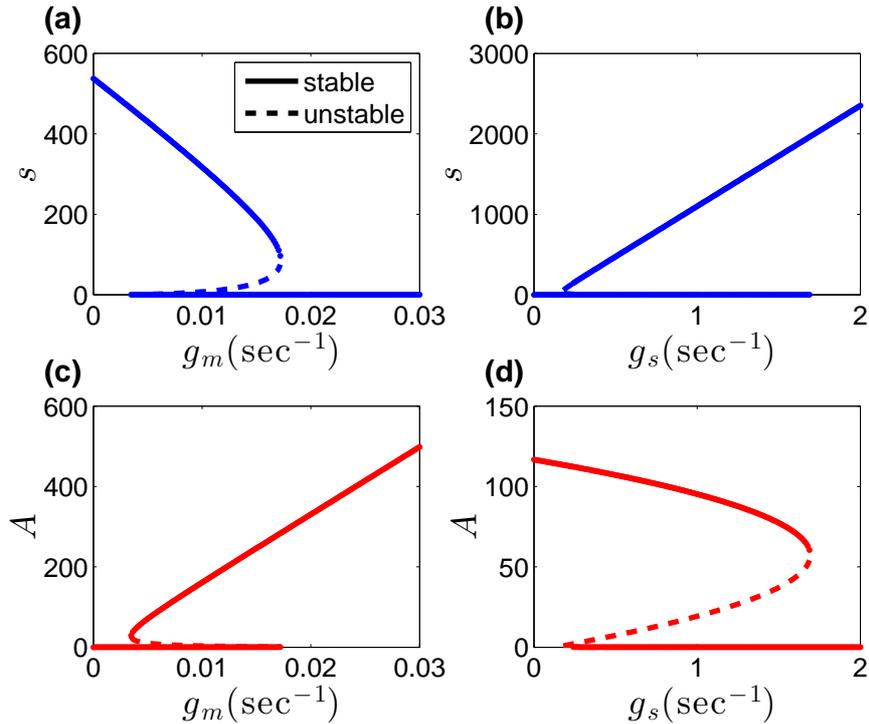}

\caption{
(Color online)
Bifurcation diagrams showing the levels of $A$ 
and $s$ as a function of the parameters $g_m$ (a,c) and 
$g_s$ (b,d). Solid lines represent stable solutions and 
dashed lines represent unstable solutions. 
}
\label{Fig_bifurcation}

\end{figure}

In Figs.
\ref{Fig_bifurcation}(a) 
and
\ref{Fig_bifurcation}(c)
we present the levels of the sRNA and the $A$ protein
under steady state conditions as a function of $g_m$,
obtained analytically from the rate equations. 
For small values of $g_m$, a single steady state is observed,
which is dominated by the sRNA.
As $g_m$ increases,
a bifurcation takes place and a second 
steady state, dominated by $A$ proteins, appears. 
A second bifurcation  
occurs at larger $g_m$ value, beyond which only
a single stable steady state remains, 
which is dominated by the $A$ proteins.
Similar results are presented in Figs. 
\ref{Fig_bifurcation}(b)
and
\ref{Fig_bifurcation}(d)
as a function of $g_s$. 

In order to extend and exemplify the robustness of the results presented above,
in Appendix B we present the bifurcation diagrams of the MFL obtained for different binding and unbinding kinetics of the TF to the sRNA promoter, a range of dissociation kinetics of the sRNA-mRNA complex, and different values of cooperativity of the TF to the sRNA promoter.
In all cases, we observe a range of parameters in which bistability takes place.


\section{Stochastic analysis}

\label{sec:stochastic}

In order to account for the effects of fluctuations 
in the MFL we analyze its dynamics using the master equation.
Here the levels of the RNAs and protein take integer values,
namely $m,A,s \in \mathbb{N}_0$, 
while the level of the bound repressor,
$r \in \{0,1\}$.
The master equation accounts for the temporal variation of 
the probability distribution 
$P({m,A,r,s})$.
It takes the form
\begin{widetext}
\begin{eqnarray}
\frac{dP({m,A,r,s})}{dt} &=& 
g_m [P({m-1,A,r,s})-P(m,A,r,s)] \nonumber \\
&+& g_s \delta_{r,0}[P({m,A,r,s-1})-P({m,A,r,s})] \nonumber \\
&+& g_A m [P({m,A-1,r,s})-P({m,A,r,s})] \nonumber \\
&+& c_{ms} [(m+1)(s+1)P({m+1,A,r,s+1})-msP({m,A,r,s})] \nonumber \\
&+& c_g [(A+1) \delta_{r,1} P({m,A+1,0,s})-A \delta_{r,0} P({m,A,0,s})] \nonumber \\
&+& u_g [\delta_{r,0} P({m,A-1,1,s})-\delta_{r,1}P({m,A,1,s})] \nonumber \\
&+& d_m [(m+1)P({m+1,A,r,s})-mP({m,A,r,s})] \nonumber \\
&+& d_A [(A+1)P({m,A+1,r,s})-AP({m,A,r,s})] \nonumber \\
&+& d_s [(s+1)P({m,A,r,s+1})-sP({m,A,r,s})].
\label{eq:sRNA_master}
\end{eqnarray}
\end{widetext}

\noindent
where $\delta_{i,j}=1$ for $i=j$ and 0 otherwise.

The first (second) term in this equation describes the 
transcription of mRNA (sRNA) molecules. 
The third term accounts for the translation of mRNAs into $A$ proteins.
The term involving $c_{ms}$ describes the binding of 
sRNA and mRNA molecules, to form an sRNA-mRNA complex.
The terms involving $c_g$ and $u_g$, 
describe the binding and dissociation of $A$ proteins to/from 
the promoter site of gene {\it b}, respectively.
The last three terms correspond to the 
degradation of mRNAs, $A$ proteins
and sRNA molecules, respectively.

In order to examine the properties of the steady state solution of the master equation, it
is useful to consider the marginal probability
distribution

\begin{equation}
P(A,s) = 
\sum_{m} 
\sum_{r} 
P({m,A,r,s}),
\label{eq:marginal_prob_dist_ms}
\end{equation}

\begin{figure}

\includegraphics[width=\columnwidth]{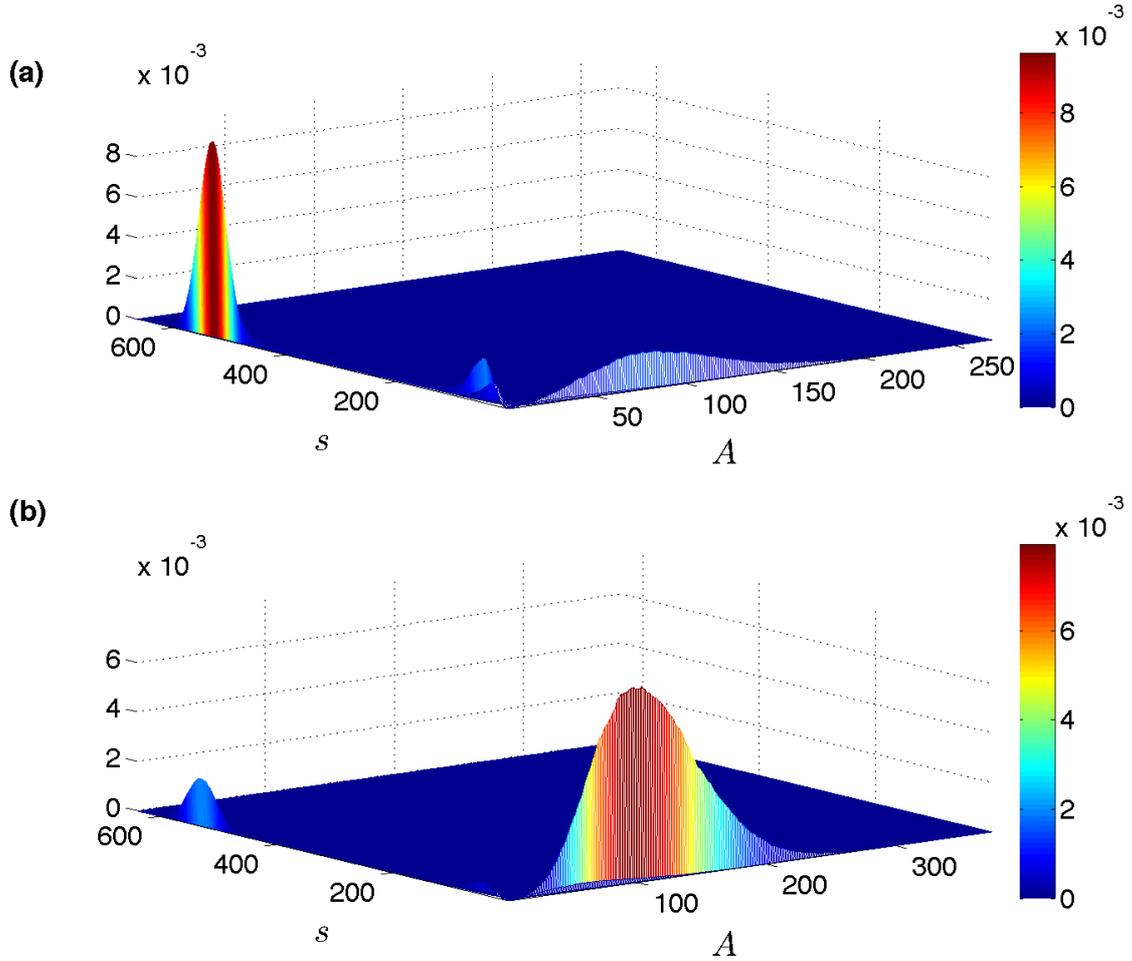}

\caption{
(Color online)
The probability distribution $P(A,s)$ for the MFL: 
(a) in the state dominated by the sRNA regulator 
(obtained for $g_m = 0.0058$ sec$^{-1})$; 
and (b) for the
state dominated by the transcriptional repressor
(obtained for $g_m = 0.0071$ sec$^{-1})$. 
}

\label{Fig_master_dist}

\end{figure}

In the formulation based on the master equation,
the criterion for bistability
is that the steady state solution 
$P(A,s)$
exhibits two distinct peaks, 
separated by a gap in which the probabilities
are low.
The locations of these peaks on the 
$(A,s)$ plane correspond to the two bistable solutions 
of the rate equations.

In order to obtain the switching times between the two bistable states and estimate the probability distributions of the different possible discrete states of the system,
we perform Monte Carlo (MC) simulations using the Gillespie algorithm
\cite{Gillespie1977}.
This is a kinetic MC
approach, namely, an algorithm that generates 'paths' of the
stochastic process. At each time
step the next move is drawn from all possible processes that
may take place at that point, where each step is endowed
with a suitable weight. After each move, the elapsed time is
properly advanced, the list of available processes is updated
and their new rates are evaluated.

In Fig. 
\ref{Fig_master_dist}
we present the probability distribution
$P(A,s)$, generated by performing MC simulations ($10^7$ sec each),
and quantifying the relative fraction of time in which the system is found in each discrete $P(A,s)$ state, averaged over initiations of the system at both the sRNA and TF dominated states.
The probability distribution is presented
for conditions under which the system is dominated
by the sRNA regulator 
[Fig. \ref{Fig_master_dist}(a)]
and for conditions under which it is 
dominated by the transcriptional repressor
[Fig. \ref{Fig_master_dist}(b)]. 
In both cases the distribution exhibits two peaks representing 
the two bistable states. In the former case the peak dominated 
by sRNAs is large and the peak dominated by transcriptional repressors
is small, while in the latter case the situation is reversed.
The peak dominated by sRNAs is sharp and narrow while the peak
dominated by the transcriptional repressors is broad.
The volume of each peak represents the cumulative probability of 
microscopic states associated with the corresponding state
of the system.
It also represents the fraction of the time in which the system is 
expected to reside in that state.

We denote the mean lifetimes of the bistable states dominated by
the sRNA and by the transcriptional repressor by
$\tau_s$ and $\tau_A$, respectively. 
To obtain the values of $\tau_s$ ($\tau_A$) we initialize
the system in the state dominated by the sRNA (transcriptional repressor)
and evaluate the 
average time elapsed until a transition to the $A$ ($s$) dominated state 
has occurred. 
The transition between states is defined
as the point 
in which the level of the previous minority species exceeds that of the
dominant species.

\begin{figure}[H]

\includegraphics[width=\columnwidth]{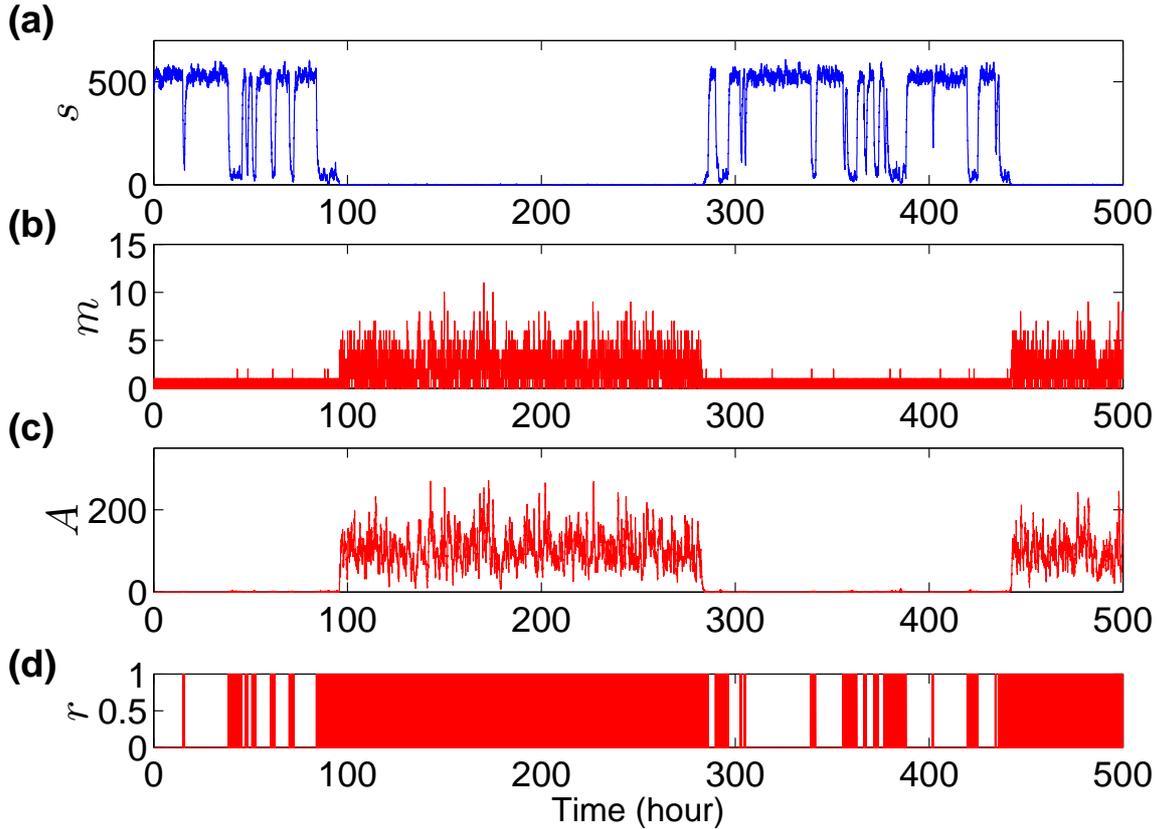}

\caption{
(Color online)
The levels of the sRNAs (a), mRNAs (b),
free $A$ repressors (c) and bound repressors (d)  
vs. time in the MFL,
obtained from Monte Carlo simulations.
Failed attempts to transition from the $s$ to the $A$ dominated state are observed, following periods in which the $s$ promoter is occupied by the transcriptional repressor.
}

\label{Fig_stoch_traj}

\end{figure}

In Fig. \ref{Fig_stoch_traj}
we present a typical
result of an MC simulation of the MFL.
The system is clearly bistable, with spontanous fluctuation-driven
switching transitions. In the sRNA dominated state there are failed
switching attempts, corresponding to the third, unstable steady state, in which the sRNA level is reduced, but is then recovered.
In the $A$ dominated state, both the mRNA and protein levels exhibit large
fluctuations, accompanied by fast binding/unbinding of $A$ proteins
to/from the {\it b} promoter.

\begin{figure}[H]
\begin{center}
\includegraphics[width=0.8\columnwidth]{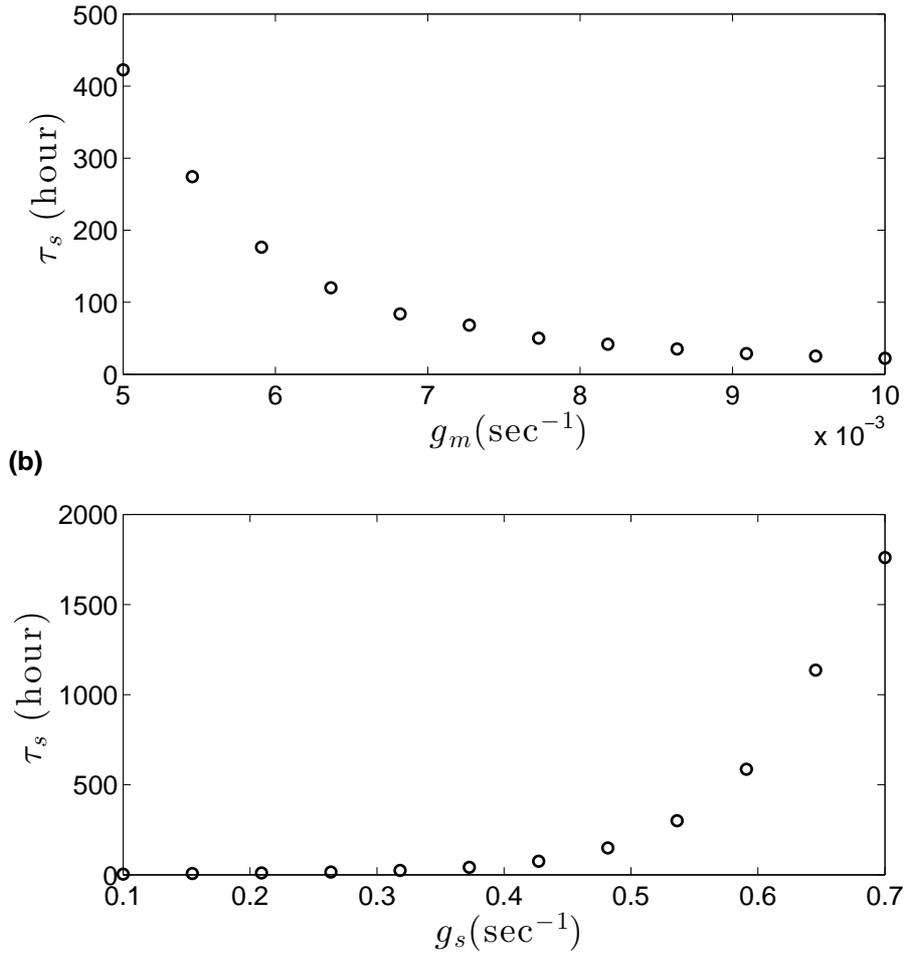}

\caption{
The average lifetime of the sRNA dominated state,
$\tau_s$, as a function of the parameters 
$g_m$ (a) and $g_s$ (b), 
obtained from MC simulations.
The lifetime $\tau_s$ monotonically decreases with $g_m$ and
monotonically increases with $g_s$. 
Each data point was averaged over 1000 MC runs.
}

\label{transition_time}
\end{center}
\end{figure}

To examine the dependence of the lifetimes of the two bistable states on parameters, we present 
in Fig. \ref{transition_time} 
the lifetime of the sRNA dominated state,
$\tau_s$,
as a function of the transcription rates of 
the mRNA and sRNA. 
As the transcription rate $g_m$, is increased,
the switching rate from the state dominated by the sRNA to the
state dominated by the transcriptional repressor increases and 
the lifetime, $\tau_s$, of the sRNA dominated state decreases 
[Fig.~\ref{transition_time}(a)]. 
On the other hand, when the transcription rate of the sRNA, 
$g_s$, is increased, the lifetime of the sRNA dominated state, 
$\tau_s$, increases [Fig.~\ref{transition_time}(b)].

Further insight into the balance between the two bistable states 
can be obtained by evaluating
the fraction of the time in which the 
system resides in each state.
The fraction of time in which the sRNA is dominant is given by

\noindent
\begin{eqnarray}
P_{s} &=& \frac{\tau_{s}}{\tau_A + \tau_s},
\end{eqnarray}

\noindent
while the fraction of time the transcriptional repressor is dominant
is
$P_{A} = 1 - P_{s}$.

\begin{figure}[H]
\begin{center}
\includegraphics[width=0.7\columnwidth]{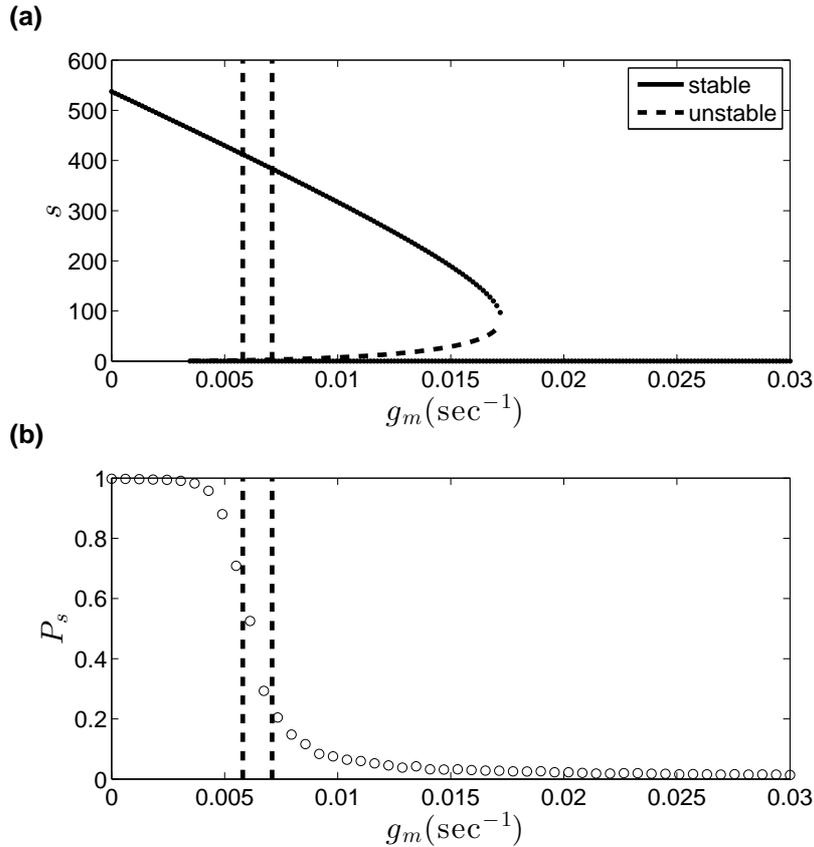}
\caption{
(a) The sRNA level, $s$, as a function of the transcription rate $g_m$,
obtained from the rate equations.
(b) The fraction of time, $P_s$, that 
the system resides in the sRNA dominated state
vs. $g_m$, obtained from MC simulations. The dependence of $P_s$ on $g_m$
is of the form of a decreasing sigmoid function.
Two vertical lines mark the $g_m$ values, as shown in Fig. 3, for the state dominated by the sRNA regulator (obtained for $g_m = 0.0058$ sec$^{-1})$, and for the 
state dominated by the transcriptional repressor (obtained for $g_m = 0.0071$ sec$^{-1})$.
}
\label{bifu_prob}
\end{center}
\end{figure}

In Fig. \ref{bifu_prob}(a)
we present the bifurcation diagram for the sRNA level vs. $g_m$,
obtained from the rate equations,
showing the range of $g_m$ values in which bistability takes place.
In Fig. \ref{bifu_prob}(b)
we present the fraction of time, $P_s$, in which the 
system resides in the sRNA dominated state vs. $g_m$,
obtained from MC simulations.
$P_s$ follows a
decreasing sigmoid function vs. $g_m$.
In the limit in which $g_m$ is small the system spends most of its 
time in the sRNA dominated state, while in the large $g_m$ limit 
it spends most of the time in the state dominated by the transcriptional
repressor.
It is found that in both limits, the MFL is biased towards one of the steady states.

\section{Summary and Discussion}

\label{sec:summary}

We have performed deterministic and stochastic analyses
of the double-negative mixed feedback loop
involving transcriptional regulation and 
post-transcriptional regulation via 
sRNA-mRNA interaction.
Using deterministic methods, 
we identified the range of parameters in which these 
systems exhibit bistability.
We have shown that within this range, 
the relative lifetimes of the two stable states ($P_s$ and $P_A$) follow complementary 
sigmoid functions as parameters such as the transcription rates
are varied.
Different $P_s$ values may be beneficial for MFLs, or other bistable systems, under different biological contexts, yielding, at the population level, a bimodal distribution.
Indeed, the relative lifetimes of biological bistable systems were extensively studied and were shown (both theoretically and experimentally) to exhibit a wide range of values \cite{acar2005,mirouze2012,santillan2008,Warren2004,Warren2005,maamar2007,isaacs2003}.

An important example of a double-negative MFL
appears in {\it E. coli}, consisting of the genes 
{\it fur} (encoding a transcriptional repressor) 
and {\it ryhB} (encoding an sRNA).
In presence of iron, the Fur repressor is active, 
repressing the transcription of the RyhB sRNA, as well as 
other genes involved in iron metabolism 
\cite{Masse2005,Masse2007,Vecercek2007}.
Thus, {\it fur} plays the role of gene {\it a}
in Fig.~\ref{fig1}.
When iron supply is limited, Fur becomes inactive and RyhB is transcribed.
Fur synthesis is translationally coupled to that of 
an upstream open reading frame, whose translation is 
downregulated by RyhB 
\cite{Vecercek2007}.
Therefore, 
{\it ryhB} plays the role of gene {\it b}
in Fig.~\ref{fig1}.
When the iron level increases and the stress condition is removed, 
the level of Fur is restored and overrides the RyhB sRNA
\cite{Masse2007}.
Another relevant example
of a double-negative
MFL in {\it E. coli} consists of the global
transcriptional regulator Lrp and the sRNA MicF
\cite{Holmqvist2012}.
Lrp activates genes that need to be expressed under nutrient-poor conditions while repressing genes that need to be expressed under nutrient-rich conditions.
Accordingly, it was shown that Lrp is highly expressed
under nutrient-poor conditions, while MicF is highly 
expressed under nutrient rich conditions.
The results presented in this paper shed further light on the sensitivity
of the dynamics of such MFLs to changes in effective parameters by 
external conditions, determining which of the two regulators dominates
under given conditions, and to what extent. 

\section{acknowledgements}

This work was supported by grants from Israel Science Foundation and  
the Israel Ministry of Science and Technology
granted to HM. MN is grateful to the Azrieli 
Foundation for the award of an Azrieli Fellowship.


\appendix


\section{Rate coefficients}

\label{sec:parameters}

The equations that describe the MFL include a large number of rate
coefficients for the rates of the transcription, 
translation, binding, unbinding and degradation processes.
To obtain results and predictions that are biologically meaningful,
one should use rate coefficients that are in the
biologically relevant range. 
While the analysis performed in this paper is quantitative, the
conclusions are of a qualitative nature and describe the generic
behavior of the MFL.
Below we discuss in more detail the considerations we have made
in order to identify the biologically relevant range of each parameter.

\subsection{Transcription and degradation rates of mRNAs}

Previous analyses have revealed that the rate limiting step in the
transcription is usually the delay between the binding of RNA polymerase
to the promoter site and the beginning of the elongation process
\cite{Mcclure1980}. 
Measurements have shown that this time lag exhibits great variation
between different genes and under different conditions, and takes values
between 20 seconds and 10 minutes.
The delay time can be represented by the transcription initiation rate,
taking values in the range
$0.001 \le g_m \le 0.05$ 
molecules per second.
Recent measurements of mRNAs in single cells showed that
for the gene that was studied, an mRNA molecule is produced 
every 7 minutes, which amounts to a transcription 
rate of $g_m = 0.0024$ (sec$^{-1}$) 
\cite{Golding2005}.
The half life of mRNA is typically in the range 
between 30 and 300 (sec).
This yields mRNA degradation rates in the range
$0.003 \le d_m \le 0.03$ ($\sec^{-1}$).

\subsection{Transcription and degradation rates of sRNAs}

An example of a small RNA 
in {\it E. coli},
on which extensive
experimental measurements were performed is OxyS.
This sRNA appears in high copy numbers 
\cite{Altuvia1997}.
In the absence of target mRNAs,
it was found to have a half-life of 12-15 minutes,
which is longer than most mRNAs 
(with typical half-life of 2-4 min).
In order to allow for variations between different sRNAs, we choose 
a broader range of half life values, translating to sRNA degradation rates in the range
$0.0002 \le d_s \le 0.002$ 
($\sec^{-1}$).

Values reported for generation rates of various sRNAs were in the range 
$0.02 \le g_s \le 7.5$ 
($\sec^{-1}$).
The lower limit was reported in \cite{Levine2007}.
The upper Limit was obtained by assuming that the steady state level of oxyS was due to synthesis and degradation processes, 
$g_s=s \times d_s = 4500 \times (0.0005-0.00167  \ \sec^{-1})$ \cite{Altuvia1997} $= 2.5-7.5 \ \sec^{-1}$.

\subsection{Protein synthesis and degradation rates}

Measurements of protein synthesis rates are reported in Ref.
\cite{Kennell1977}.
It was shown that a protein can be translated from each mRNA molecule 
every 3-4 seconds.
To cover a broad range of biologically relevant translation rates,
we take the range of 2-20 seconds.
This corresponds to translation rate coefficients in the range
$0.05 \le g_A \le 0.5$ (sec$^{-1}$).

Transcription factors are usually short lived, 
with half life of a few minutes. 
Thus, we consider degradation rates in the range
$0.001 \le d_A \le 0.05$ (sec$^{-1}$).

\subsection{Binding rates of sRNAs to their mRNA targets}

Measurements of the binding of the sRNA OxyS
to its mRNA target {\it fhlA}, performed in vitro,
are reported in Ref.
\cite{Altuvia1997}.
In these experiments
2nM of the OxyS sRNA were mixed with different 
concentrations of the {\it fhlA} mRNA.
After 5 minutes the concentration of free OxyS was measured.
It was found that when the concentration of the 
{\it fhlA} mRNA was 25nM, 
half of the OxyS molecules were bound after 5 minutes.
This was done in vitro, where the synthesis of new sRNA and
mRNA molecules as well as their 
degradation were suppressed.
Denoting the level of OxyS by $s$ and of the {\it fhla} mRNA
by $m$, the dynamics can be described by
\begin{eqnarray}
\frac{ds}{dt} &=& - c_{ms} m \cdot s \nonumber \\
\frac{dm}{dt} &=& - c_{ms} m \cdot s.
\end{eqnarray}

\noindent
This means that under these conditions 
the difference between the levels of $m$ and $s$ remains constant. 
Assuming that the initial levels at time $t=0$ 
satisfy
$m_0 > s_0$,
we can solve this equation and obtain
\begin{eqnarray}
s(t) &=& \frac{(m_0-s_0) s_0}
{m_0 \exp[c_{ms} (m_0-s_0)t] - s_0}  \nonumber \\
m(t) &=& \frac{(m_0-s_0)m_0}{m_0-s_0\exp[-c_{ms}(m_0-s_0)t]}.
\end{eqnarray}

\noindent
Setting the initial conditions 
and fitting the binding rate coefficient
$c_{ms}$ such that 
after 5 minutes the sRNA concentration $s$
goes down to a half of the initial concentration $s_0$,
we obtain that
$c_{ms}=9.45 \times 10^{-5} ({\rm nM}\cdot\sec)^{-1}$.
Taking the {\it E. coli} cell volume as $10^{-15}$ liters, 
we obtain 1nM=0.6 molecules per cell, giving
$c_{ms}=0.0002$ 
(sec$^{-1}$).
Since this experiment was carried out without Hfq, 
which is a catalyst of the reaction,
it is reasonable to take a range 
of $c_{ms}$ values
which express faster binding.
This would most likely also account for variations in the binding rates
of other sRNA molecules to other mRNAs.
We therefore take the range
$0.001 \le c_{ms} \le 0.02$ 
(sec$^{-1}$).

\subsection{Transcription factor binding/unbinding rates}

The binding and unbinding rates of two transcription factors
in two {\it E. coli} strains to/from their specific promoter sites
on the DNA were measured in Ref.
\cite{Henriksson2007}. 
Using surface plasmon resonance, which can monitor the 
time dependent changes in concentrations,
they found binding rates of $0.09 - 0.14$ 
(sec$^{-1}$).
This means that a transcription factor would bind to the DNA within
7-11 seconds.
Measuring the ratio between bound and free DNA 
yields the ratio between 
the binding and dissociation rates.
The values that were obtained for the dissociation rate are 
in the range of
$0.001 - 0.002$ 
(sec$^{-1}$).
This in turn means that a transcription factor stays bound to the
promoter site for 1000 to 2000 seconds.
To make the range more dynamic and account for weaker
transcriptional repression we choose the ranges
$0.05 \le c_g \le 0.25$ (sec$^{-1}$),
and
$0.001 \le u_g \le 0.01$ 
(sec$^{-1}$).


\section{Extended stability analysis}
\label{stability_range}

Here we investigate how the MFL bifurcation diagram is affected by a broad family of parameter variations and modifications of the network.
In Fig. 2 we showed the bifurcation diagrams for the parameters $g_m $ and $g_s$, representing the transcription rates of the TF and sRNA, respectively.
Here we further present the bifurcation diagrams for 
the binding and unbinding rates of the TF to the sRNA promoter, denoted by $c_g$ and $u_g$, respectively,
and the dissociation rate of the sRNA-mRNA complex, denoted by $u_c$.
In addition, we examine the effect of cooperativity of the binding of the TF, $A$, to the sRNA promoter which is characterised by the Hill coefficient $\alpha$.
The rate equations describing the MFL with cooperative binding take the form

\begin{subequations}
\label{eq-alpha}
\begin{align}
\frac{dm}{dt} &= g_m - d_m m - c_{ms} m \cdot s + u_c C \\
\frac{ds}{dt} &= g_s (1-r) -d_s s - c_{ms} m \cdot s + u_c C \\
\frac{dA}{dt} &= g_A m - d_A A -\alpha c_g A^{\alpha}(1-r) +\alpha u_g r \\
\frac{dr}{dt} &= c_g A^{\alpha}(1-r) - u_g r \\
\frac{dC}{dt} &= c_{ms} m \cdot s - d_c C - u_c C \ .
\end{align}
\end{subequations}

\begin{figure}[H]

\includegraphics[width=0.95\textwidth]{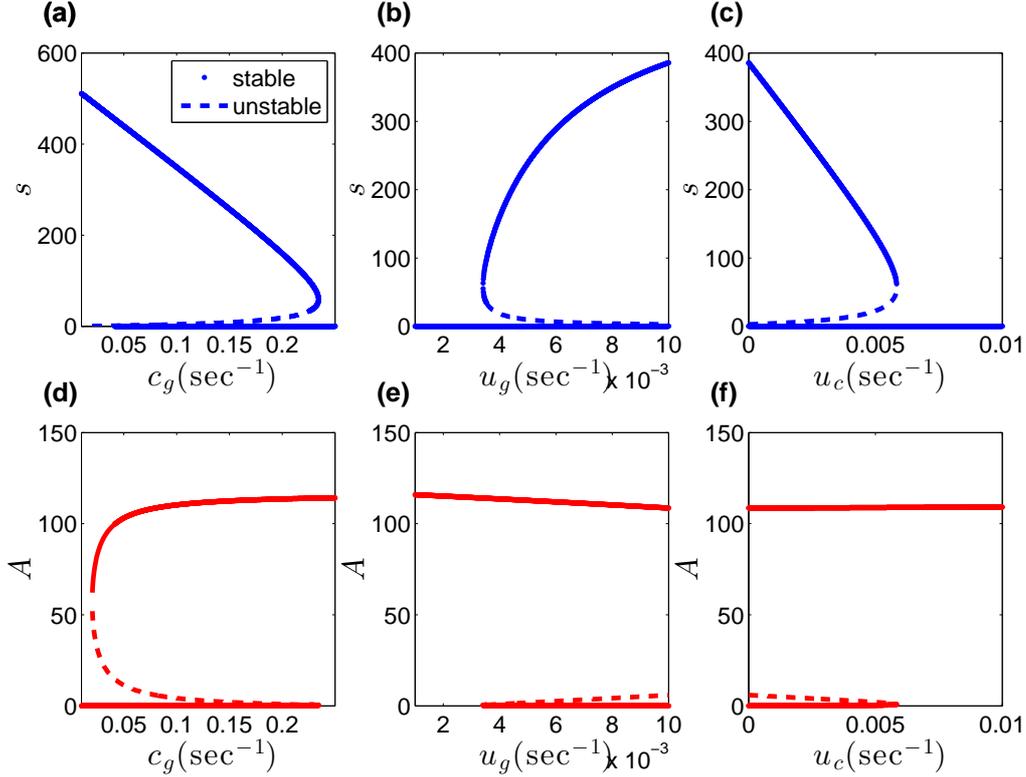}

\caption{
(Color online)
Bifurcation diagrams showing the levels of $A$ 
and $s$ as a function of the parameters $c_g$ (a,d), 
$u_g$ (b,e), and $u_c$ (c,f). Solid lines represent stable solutions and 
dashed lines represent unstable solutions. 
}
\label{Fig2_extended}

\end{figure}

\noindent
For the results presented below, we take the following default parameter values:
$g_m$=0.007, $g_s$=0.43, $g_A$=0.05, 
$d_m$=0.003, $d_s$=0.0008, $d_A$=0.001, 
$c_{ms}$=0.02, $c_g$=0.08, $u_g$=0.01,
$d_c$=0.003, $\alpha$=1.
All the parameters are in units of sec$^{-1}$, except for $\alpha$ which is dimensionless.
In Figs.
\ref{Fig2_extended}(a,b) 
and
\ref{Fig2_extended}(d,e)
we present the levels of the sRNA and the $A$ protein
under steady state conditions as a function of $c_g$ and $u_g$, respectively,
obtained analytically from the rate equations. 
The ranges of $c_g$ and $u_g$ agree with those presented in Appendix \ref{sec:parameters}.
As expected, the effect of $c_g$ (binding of the TF to the sRNA promoter) opposes that of $u_g$ (unbinding); as $c_g$ increases or $u_g$ decreases, the TF strengthens its repressive role over the sRNA. 
More specifically, 
for small values of $c_g$, a single steady state, dominated by the sRNA, is observed.
As $c_g$ increases,
a bifurcation takes place and a second 
steady state, dominated by $A$ proteins, appears. 
A second bifurcation  
occurs at larger $c_g$ value, beyond which only
a single stable steady state remains, 
which is dominated by the $A$ proteins.
For small values of $u_g$, a single steady state, dominated by the TF, is observed,
As $u_g$ increases, a bifurcation takes place and a second 
steady state, dominated by the sRNA, appears. 
In Figs.
\ref{Fig2_extended}(c) 
and
\ref{Fig2_extended}(f)
we present the levels of the sRNA and the $A$ protein
as a function of $u_c$.
Here, increasing $u_c$ 'weakens' the sRNA state, yielding, following a bifurcation, a single steady state, dominated by $A$.

\begin{figure}[H]

\includegraphics[width=0.95\textwidth]{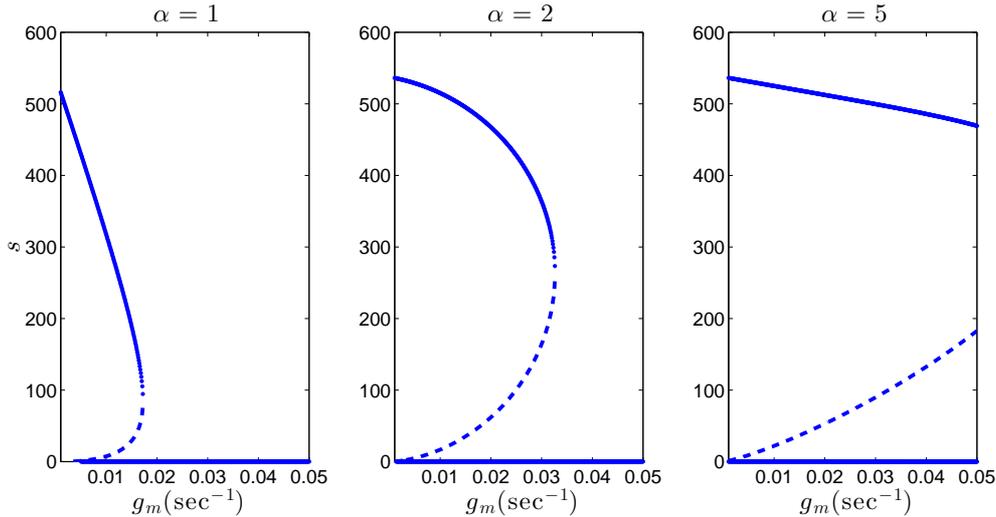}

\caption{
Bifurcation diagrams showing the level of $s$ as a function of the parameter $g_m$, for different values of $\alpha$. Solid lines represent stable solutions and 
dashed lines represent unstable solutions. 
}
\label{Fig_alpha}

\end{figure}

Next, we consider the effect of the cooperativity, as expressed by the Hill coefficient $\alpha$ in Eq. \ref{eq-alpha}, of the binding of the TF protein to the sRNA promoter on the stability.
In Fig.
\ref{Fig_alpha}
we present the bifurcation diagram showing the level of $s$ as a function of $g_m$, for different values of $\alpha$ ($\alpha=1,2,5$). 
The range of $g_m$ agrees with that presented in Appendix \ref{sec:parameters}.
As expected, as cooperativity increases, the repression by $A$ weakens and the $s$ state strengthens. This is expressed as a higher-level steady state and delayed bifurcation (in terms of $g_m$) for higher $\alpha$, as can be seen in Fig. \ref{Fig_alpha}.
Note that for $\alpha=1$, $c_g$ is simply the binding rate of the transcriptional repressor to the sRNA promoter. For $\alpha>1$, it represents the overall rate of a more complicated process, which includes the assembly of a repressor complex of $\alpha$ repressors, and its binding to the promoter site. 

\bibliographystyle{prsty}

\bibliography{mfl}

\end{document}